# Slanted light-sheet array microscopy for large volume imaging at rates exceeding 100 Hz

Kai Long,[a] Wenkai Chen,[a] Junming Zhou,[a,b] Junyi Li,[b] Shuhao Shen,[c] Zhipeng Tai,[d] Shifeng Xue,[d] Anqi Qiu,[e] and Nanguang Chen,[a,b,*]

[a]National University of Singapore, Department of Biomedical Engineering, 4 Engineering Drive 3, Singapore, 117583
[b]NUS (Suzhou) Research Institute, No. 377 Linquan Street, Suzhou Industrial Park, Suzhou, Jiangsu, China
[c]Xidian University, Guangzhou Institute of Technology, Guangzhou, China, 510555
[d]National University of Singapore, Department of Biological Sciences, 16 Science Drive 4, Singapore, 117558
[e]Hong Kong Polytechnic University, Department of Health Technology and Informatics, Hong Kong

**Abstract**. High-speed image acquisition in light microscopy is essential for a wide range of applications, including observing dynamic biological processes and enabling high-throughput sample analysis. However, traditional imaging speeds are often limited by the scanning mechanisms and the signal-to-noise ratio, and these constraints are further exacerbated by the need for volumetric imaging, optical sectioning, high spatial resolution, and large fields of view. To address these challenges, we have developed a slanted light-sheet array microscope (SLAM), which enables ultrafast volumetric imaging without compromising key technical specifications. SLAM is built on a standard wide-field compound microscope with minimal and straightforward modifications to the illumination path, allowing for easy integration. It can acquire multi-dimensional, high-resolution images at rates exceeding 100 volumes per second across large imaging regions (e.g., exceeding 500 pixels in transverse dimensions and 200 layers in depth). In addition, a deep learning approach based on conditional denoising diffusion probabilistic models is proposed to achieve isotropic resolution. Like traditional light-sheet microscopy, SLAM offers intrinsic optical sectioning and localized photochemistry, while its innovative optomechanical design is compatible with most biological samples prepared using conventional protocols. This makes SLAM a versatile and powerful imaging platform that is accessible to the broader biomedical research community.

*Correspondence to: Nanguang Chen, Email:  biecng@nus.edu.sg

## 1 Introduction

Volumetric imaging is indispensable for capturing the complexity of biological systems in three dimensions[1-6]. It enables breakthroughs in understanding fundamental biology and improving diagnostic and therapeutic strategies in clinical medicine. Biological tissues are three-dimensional in nature. Volumetric macroscopic, mesoscopic, and microscopic imaging methods allow direct observation of the intricate architecture of tissues, cells, and subcellular components in their native 3D context. It contributes to a deeper understanding of biological functions, cell interactions, and developmental processes that are often lost in two-dimensional imaging. In personalized medicine, high-resolution volumetric data supports the development of personalized treatment plans by revealing individual anatomical and pathological variations, thereby improving outcomes and reducing risks[7]. In tissue engineering, 3D imaging is crucial for characterizing scaffold structures, monitoring cell growth, and assessing tissue regeneration, ensuring the development of functional biological constructs[8-10]. Many biological phenomena are highly dynamic. Neural activity, blood flow, cell migration, and developmental changes occur in three dimensions and over time. Volumetric imaging with adequately high temporal resolution is vital for real-time observation of these dynamic events, providing insights into mechanisms that are crucial for understanding health and disease[11,12].

Over the past few decades, light microscopy has evolved from traditional compound microscopes for mechanically sliced thin samples to sophisticated, multi-dimensional imaging platforms. Laser scanning microscopes, including confocal microscopes[2,13,14] and multiphoton microscopes[15-18], provide excellent optical sectioning capabilities that enable high-quality, layer by layer visualization of cellular and sub-cellular structures. However, they usually require higher illumination power and are known for increased photobleaching and phototoxicity. In addition, the point-by-point scanning of samples slows down the image acquisition process. Typically, it takes more than a few hundred milliseconds to obtain a 2D image of 512 by 512 pixels. Line-scan confocal microscope and its variants have been demonstrated to reach a much higher imaging speed. For example, Pant et al. reported a line-scan focal modulation microscope, which captured high-contrast fluorescence images from a beating zebrafish heart at 100 frames per second[19,20]. Temporal focusing, with its ability to focus light in time, eliminates lateral scanning and leads to improved temporal resolution for nonlinear fluorescence microscopy[16,17]. Higher imaging speed can also be achieved with multifocal multiphoton microscopy[18]. Albeit all these technical advances, volumetric imaging at more than 1 Hz is rare for laser scanning microscopy methods.

Light-sheet microscopy becomes increasingly popular due to its intrinsic optical sectioning capability afforded by the use of selected plane illumination[21-23]. It allows high speed imaging as an area imaging sensor collects emission light signals from the illumined sample layer in parallel. Furthermore, photobleaching and phototoxicity are minimized as the sample regions outside the focal plane are not illuminated during image acquisition[24]. Light-sheet microscopy has been widely used in biological research involving translucent small animal models such zebrafish embryos and larvae[25-28]. It plays an important role in life science research, from developmental biology and observation of small animal models to structural analysis of cells and tissues, to fine imaging of molecular dynamics.

A conventional light-sheet microscope has two objectives, one for illumination and the other one for detection and they are arranged to be orthogonal to each other. Due to the geometrical configuration and consequential space constraints, a conventional light-sheet microscope can only accommodate specialized sample chambers and is not compatible with typical biological samples prepared with established protocols. To overcome this limitation, many light-sheet microscope variants have been developed, including single-objective light-sheet microscopes[29,30] and open-top (single-view or dual-view) light-sheet microscopes[31-35]. Among single-objective light-sheet microscopes, there are oblique plane microscope (OPM)[36], scanned oblique plane illumination (SOPi) microscope[37,38], epi-illumination SPIM (eSPIM)[39], and swept confocally aligned plane excitation (SCAPE) microscope[40,41]. All these techniques use a single, high NA objective to illuminate the sample with a scanning light-sheet and collect the emission light propagating in the normal direction. To couple the tilted illumination plane to an imaging sensor, two remote objectives of high NA are positioned in a complicated way to maintain proper and efficient light detection. The optical design of such systems is complex and requires careful optical alignment, not to mention that the use of three high-NA objectives significantly increases the lower bound of system cost[42,43].

Here we propose slanted light-sheet array microscopy (SLAM), a novel design aiming to provide a mesoscopic and microscopy platform for convenient, ultrafast, volumetric imaging of diverse biological samples ranging from flat tissue slides to live animal models. Our system deviates significantly from the orthogonal arrangement of the illumination and detection optics in a conventional light-sheet microscope. Its detection lightpath is identical to that of a typical upright compound microscope, while the condenser is replaced by a light-sheet illuminator. The light-sheet illuminator generates a light-sheet or light-sheet array, which is directed to the sample at a slanted angle. A fast-scanning mechanism shifts the light-sheet (array) along the lateral direction and rapidly scan through the sample volume. We have validated its technical advantages in terms of temporal resolution, field of view, and optical sectioning with a wide variety of biological samples.

## 2 Results

### 2.1 Slanted light-sheet array microscope

The SLAM system was built on a standard upright wide-field microscope (Nikon, model No. ECLIPSE 80i). This upright sample stage is compatible with glass slides and various types of vessels for live sample imaging (e.g. glass bottom dishes, flasks, and multi-well plates). **Fig. 1a**. illustrates the overall system architecture, highlighting the major components and the optical design. Essentially the condenser (white light illuminator) underneath the sample stage was replaced by a light-sheet array illuminator, which included a laser module iChrome CLE-50. The laser module provided a maximum output power of approximately 50 mW at the fiber tip for any of four wavelengths: 405nm, 488nm, 561nm, and 640nm. The specifications of this laser source allowed for flexible multi-wavelength excitation of most fluorescence dyes, enabling various fluorescence imaging modes. Its fiber output was collimated and then expanded for proper beam forming. A key component in the illuminator was a cylindrical lens that condensed the excitation light to an illumination plane. In an alternative and high-speed configuration, the cylindrical lens was replaced by a cylindrical microlens array (#86-843, Edmund Optics) to generate an array of light-sheets. The illuminator further included a few spherical lenses, including an illumination objective (IO), and a mirror (M1) to direct the excitation beam towards the sample stage. Upon removal M1, the imaging system is reduced to a conventional wide-field microscope.

**Fig. 1b** shows how a slanted light-sheet was formed and the scanning mechanism. Each light-sheet formed immediately after the IO propagated upwards along the direction parallel to the optic axis of the detection objective (DO). In a typical configuration, a 4X Olympus objective (NA 0.1) was used as the IO, where the thickness of each light-sheet was about 3 μm. The propagation direction of the light-sheet (array) was bent for typically 30 to 60 degrees after it passed through a transmission grating (TG) placed between the IO and the sample stage. The 1st order diffraction resulted in a slanted light-sheet (array) for sample illumination. For optimal light-sheet generation, we selected a custom designed volume phase holographic (VPH) transmission grating (Wasatch Photonics) to enhance the 1st order diffraction efficiency (87.4%) and minimize other orders. It was positioned closely to the glass slide underneath a sample or the glass bottom dish mounted on the stage. The small gap between the VPH and the microscope slide bottom surface was filled with a matching medium, which allowed proper illumination light coupling and free translation of the sample stage relative to the illumination beam. Sample scanning was achieved by using a galvo-mirror (Thorlabs, model No. GVS011), which was responsible for shifting each light-sheet laterally at a repetition rate greater than 200 Hz. The detection objective (DO) above the sample stage could collect both scattering light and fluorescence emissions from the sample illuminated by the light-sheet array. In case of fluorescence imaging, a fluorescence filter set was inserted into the detection light path to reject the excitation light. Raw images were captured by a high-speed scientific CMOS (sCMOS) camera (pco.dimax cs1, Excelitas Technologies® Corp), which offered fast frame rates of up to 3086 fps at a full resolution of 1296 x 1024 pixels.

The raw image acquisition process was demonstrated with a zebrafish larva (3 dpf), for which the vasculature was fluorescence labelled (see **Methods**, Sample preparation procedures). As illustrated in **Fig. 1c**, at each step in the scanning process, a few slanted layers inside the sample volume were illuminated by the light-sheets at discrete lateral positions, which were evenly spaced along the lateral scanning direction (typical distance ~100 μm). Structural/functional information from illuminated planes was projected simultaneously onto the image sensor, as indicated by the spatially separated yellow rectangles (dashed line) in raw image frames (**c1**). As the light-sheet array shifted step by step (**c2**), raw image frames were obtained sequentially until the full volume of interest was covered. The total number of scanning steps was significantly reduced due to parallel illumination with the light-sheet array, as the lateral shift only needed to cover a small distance of 100 μm.

To visualize the entire sample volume, a reconstruction algorithm was then applied to the raw images to generate a 3D sample image stack. 3D image reconstruction was a straightforward reverse mapping. It involved assigning pixels in each raw 2D frame to corresponding voxels in the reconstructed 3D volume (**c3**). Interpolation/extrapolation could be employed to achieve optimal image quality. Shown in **Fig. 1d** is the z-projection of the reconstructed SLAM image volume for the entire zebrafish larva. A 3D view of a middle trunk region (marked by the yellow rectangle) is rendered in **Fig. 1e**.

It should be noted that while the current SLAM system was built on top of an upright compound microscope, it is straightforward to adapt the design to inverted microscopes.

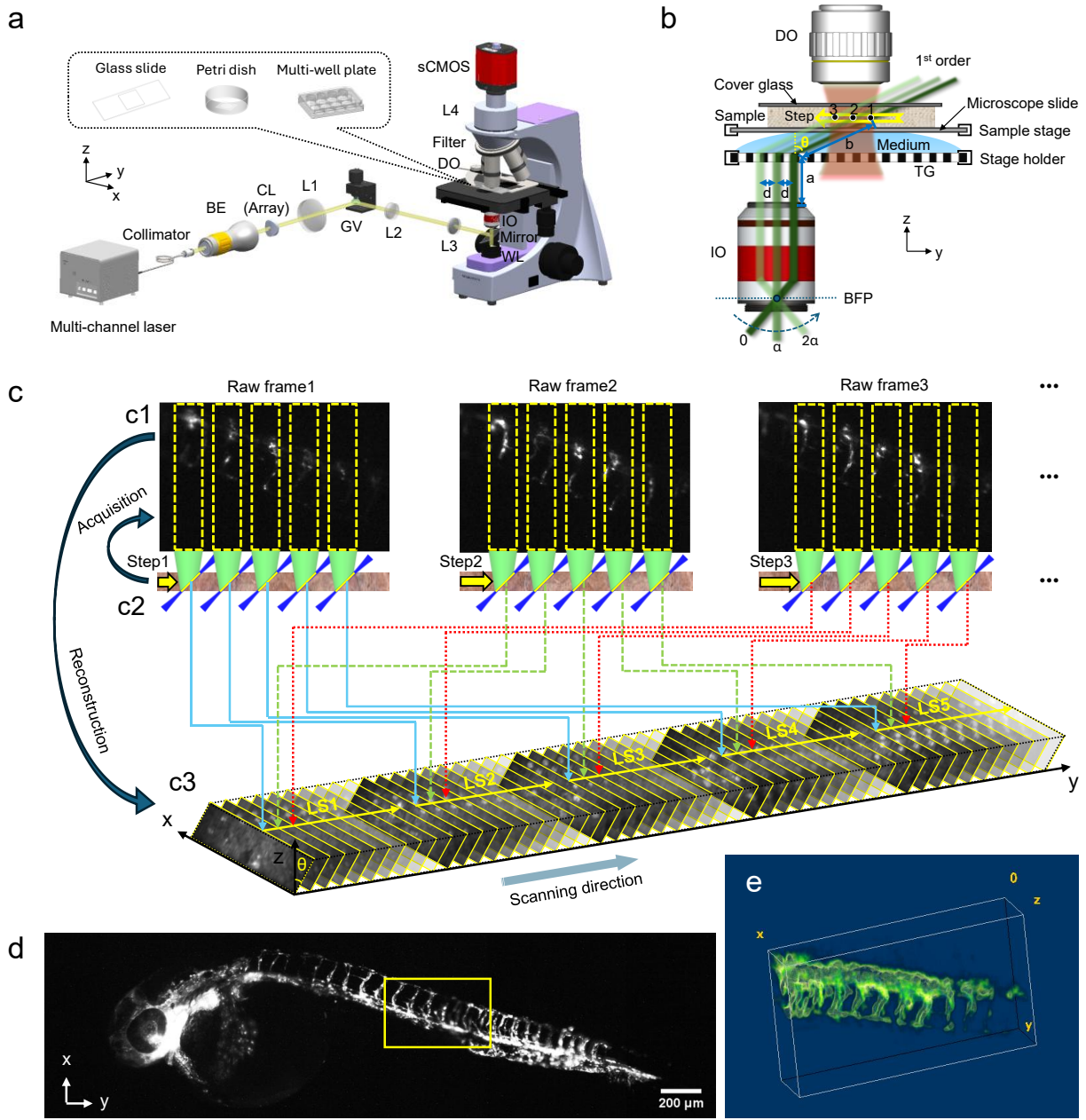

**Figure 1| SLAM design and principle.** (**a**) Schematic of a slanted light sheet array microscope. BE, beam expander; CL, cylindrical lens; GM, galvo-mirror; L1, L2, L3, and L4 lenses; IO, illumination objective; DO, detection objective; WL, white light illuminator. (**b**) Light sheet tilting and scanning. TG, transmission grating; BFP, back focal plane of IO. The 1$^{st}$ order diffraction of each light-sheet by the TG provides slanted plane illumination (with a tilting angle $\theta$), which shifts from right to left step by step at an incremental distance of $d$ in repones to the increasing GM scanning angle. The sum of a and b is around the working distance of IO. (**c**) Raw image formation and 3D image reconstruction. **c**1, raw images formed at three scanning steps; **c**2, cross-sectional view of the sample illuminated by the light-sheet array; **c**3, volume view of an image stack reconstructed from raw images. LS1, LS2, … LS5, light-sheet No. 1, light-sheet No. 2, …, light-sheet No. 5. (**d**) Z-Projection of 3D fluorescence image of the zebrafish larva whole vasculature. (**e**) The 3D rendering of rectangle section in (**d**).

## 2.2 *Multi-dimensional imaging of zebrafish larva in vivo*

SLAM provides an exceptional scanning mechanism, which eliminates the imaging speed bottleneck in most microscopic methods. In this section, we demonstrate the superior speed of SLAM in live small animal imaging.

Scattering imaging is based on an intrinsic contrast mechanism. It can provide label-free, morphological and/or functional (e.g., blood flow) images of the sample. SLAM is especially suitable for scattering based imaging as the angle between the incident light-sheet and the scattered light towards the DO is typically very large. Due to the strong forwarding scattering nature of most biological soft tissues, the scattering signal collected by the DO is typically so strong that an exposure time as small as a few microseconds would be adequate[44]. Therefore, the imaging speed is not limited by the signal to noise ratio even for a frame rate beyond 100,000 Hz.

We conducted SLAM scattering imaging experiments with a zebrafish larva (see **Methods** for more information). The live small animal sample was illuminated using a 640 nm laser. We chose an Olympus DO (20x, NA 0.4) and an IO with a NA around 0.1. This combination provided both high resolution and an appropriate depth of field (DOF). The GM was driven at 100 Hz with a sawtooth waveform. The camera's FOV was set to 285 μm × 267 μm (576 pixels × 540 pixels) and the frame rate was 5500 frames per second. The scanning step size was 2 μm, while 55 frames were acquired to complete a volume scan. The transverse scanning range was slightly greater than the lateral distance between neighboring light sheets (~100 μm). Therefore, 51 out of every 55 frames were selected to reconstruct one 3D volume whiles the redundant frames were discarded. This setup allowed us to acquire 100 complete volumes in one second, which was adequately fast to visualize the beating heart. Bearing in mind, however, this is not the maximum capability of the system.

The reconstructed image stack was a 4D hyperstack (3D in space and 1D in time). There were 100 time points evenly spread over a one-second period. At each time point, the 3D volume consisted of 245 *en face* images of 576 pixels × 540 pixels. The voxel size was 0.5 μm × 0.5 μm × 0.5 μm. **Fig. 2a** presents orthogonal views (**a1**, **a2**, and **a3**) of the imaging volume (at the time point No. 15), in which the beating heart could be well recognized from all aspects. A Z-projection was created and provided in **a4** for comparison. We moved on to retrieve functional information from the image stack. Particle Imaging Velocimetry (PIV) and laser speckle contrast analysis are established methods for generating flow maps from light scattering images[44-46]. **Fig. 2b** shows example PIV analysis results, which are vectorial velocity maps in both transverse (**b1**) and cross-sectional (**b2**) planes. Shown in **Fig. 2c** are velocity maps derived from laser speckle contrast estimated temporally and spatially. Panel **c1** is an instantaneous velocity map, which was comparable with **b1** expect that no directional information was available. Both analyses methods suggested that the motion (including blood flow and bulk tissue movement) was primarily confined to the heart region. Panels **c2** and **c3** are 3D projections of the volumetric velocity distributions at two different time points. The rotation angles were evenly distributed in the 360º range. Three locations inside the heart region (marked by small color squares) were picked to trace the temporal changes in local velocities and the results compared in **Fig. 3d** showed apparent amplitude and phase differences. It is evident that the high volumetric imaging speed could lead to rich information and comprehensive analysis of the sample under investigation.

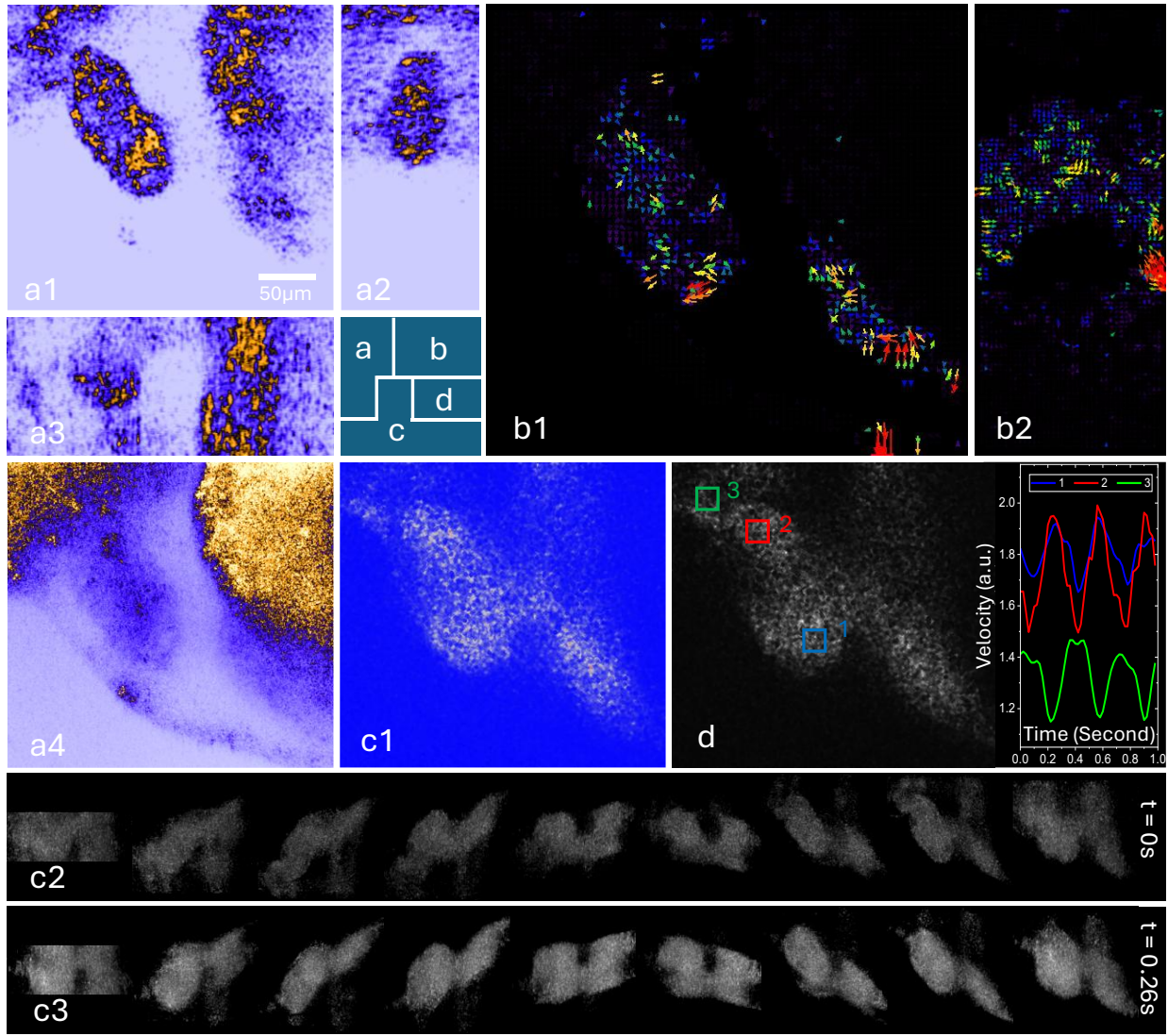

**Figure 2 |** *In vivo* **imaging of zebrafish larva at 100 volumes per second.** (**a**) Orthogonal views of the heart at $t =$ 140 ms. **a1**, en face image; **a2** and **a3**, cross-sectional views; **a4**, z-projection. (**b**) PIV generated vectorial velocity maps in transverse (**b1**) and cross-sectional (**b2**) planes. (**c**) Flow maps estimated from laser speckle contrast. **c1**, an

instantaneous flow map in the XY plane; **c2**, 3D projection of the volumetric flow distribution at $t = 0$. The rotation angle was 40 degrees per step; **c3**, 3D projection of the volumetric flow distribution at another time point $t = 0.26s$ for comparison. (**d**) Dynamic flow velocity changes measured at different locations. The waveforms and measurement locations (squares in the flow map) were color matched.

## *2.3 Fast scanning of flat biological samples*

Most biological samples are flat. Their lateral dimensions are usually far greater than the thickness. Examples of flat samples include prepared microscope slides (biology, botany, zoology, pathology, histology, hematology, parasitology, genetics, etc.), microfluidic devices, and live cells in culture dishes. Conventional light-sheet microscopes are not compatible with flat samples, while laser scanning microscopes are frequently utilized to obtain 3D images from them.

SLAM is an excellent alternative imaging technique for volumetric visualization of flat biological samples. It can scan the samples with significantly reduced imaging time and photochemistry. In this section, we demonstrate the application of SLAM with two different tissue samples: one mouse lymph nodes slide and one mouse intestine section.

Branchial lymph nodes were harvested from WT and ApoE-/- mice and cryo-sectioned at 100 μm. T cells and B cells were surface labelled with CD3 in Cy3 channel and B220 in Cy2 channel respectively. The sections were counterstained with DAPI for cell nuclei visualization. While the tissue section was about 100 μm thick, the fluorescence signal was confined to a smaller depth range probably due to the limited penetration of fluorescence labels. **Fig. 3a** shows color merged SLAM images of the sample, including a volume view and *en face* images at different depths. Fluorescence labels Cy3, Cy2, and DAPI were rendered in red, green, and blue, respectively. The stained structures varied remarkably from one layer to another even though the depth incremental was merely 2 μm, suggesting a good optical sectioning capability. The SLAM stack was acquired with a 40x Zessis objective (NA~1.1) and consisted of 150 reconstructed fluorescence images (1296 x 1024 pixels) in a depth range of 37.5 μm. For each of three fluorescence channels, the scanning time was 16 seconds as we used a very small step size of 0.25 μm to ensure oversampling. If the same region of interest were scanned with a confocal microscope with a pixel dwell time of 10 μs, the image acquisition time would be 398 seconds to obtain 30 depth layers. Besides other benefits, the fast imaging process would significantly reduce the sample exposure to excitation light and consequent photobleaching.

A FluoTissue 550-μm-thick Mouse Intestine Section was purchased from SunJin Lab Co. (Specimen No. PS002). The vasculature of the mouse intestine section was labeled with Alexa Fluor® 488, while the cell nuclei were stained using SYTOX® Orange. The sample was first scanned with a mesoscopic imaging configuration, which involved a low NA (~0.1) Olympus 4x objective lens for fluorescence emission collection. **Fig. 3b** presents orthogonal views, revealing the three-dimensional architecture of the intestinal tissue. Even though the detection NA was rather low, linear vascular features with widths of less than 10 μm could be clearly resolved in the cross-sectional views (**b2** and **b3**). Well-defined circular structures marked in all three orthogonal planes were approximately 40 μm in diameter. It took 40 seconds to complete the scan of the total imaging volume of 2.592 mm x 2.048 mm x 0.5 mm for one excitation wavelength, while the voxel size was 2 μm in all three dimensions. The same sample was then imaged with a more microscopic SLAM configuration for higher resolutions. We used an Olympus 10x objective (NA~0.25) for fluorescence detection and a NA0.3 IO to create thin illumination light-sheets. Consequently, the field of view was reduced to 648 μm x 512 μm and the depth range decreased more significantly to around 42 μm. The voxel size, however, became much smaller at 0.5 μm. Shown in **Fig. 3c** are vasculature images (Alexa Fluor® 488). While **c1** was an *en face* image for the full FOV, **c2** was a magnified image enclosed by the yellow rectangle and **c3** was a 'Slice & Borders' view (created with ImageJ 3D viewer) of another subregion marked by the green box. The image stack was acquired in 8 seconds as about 10 light-sheets were involved for parallel illumination.

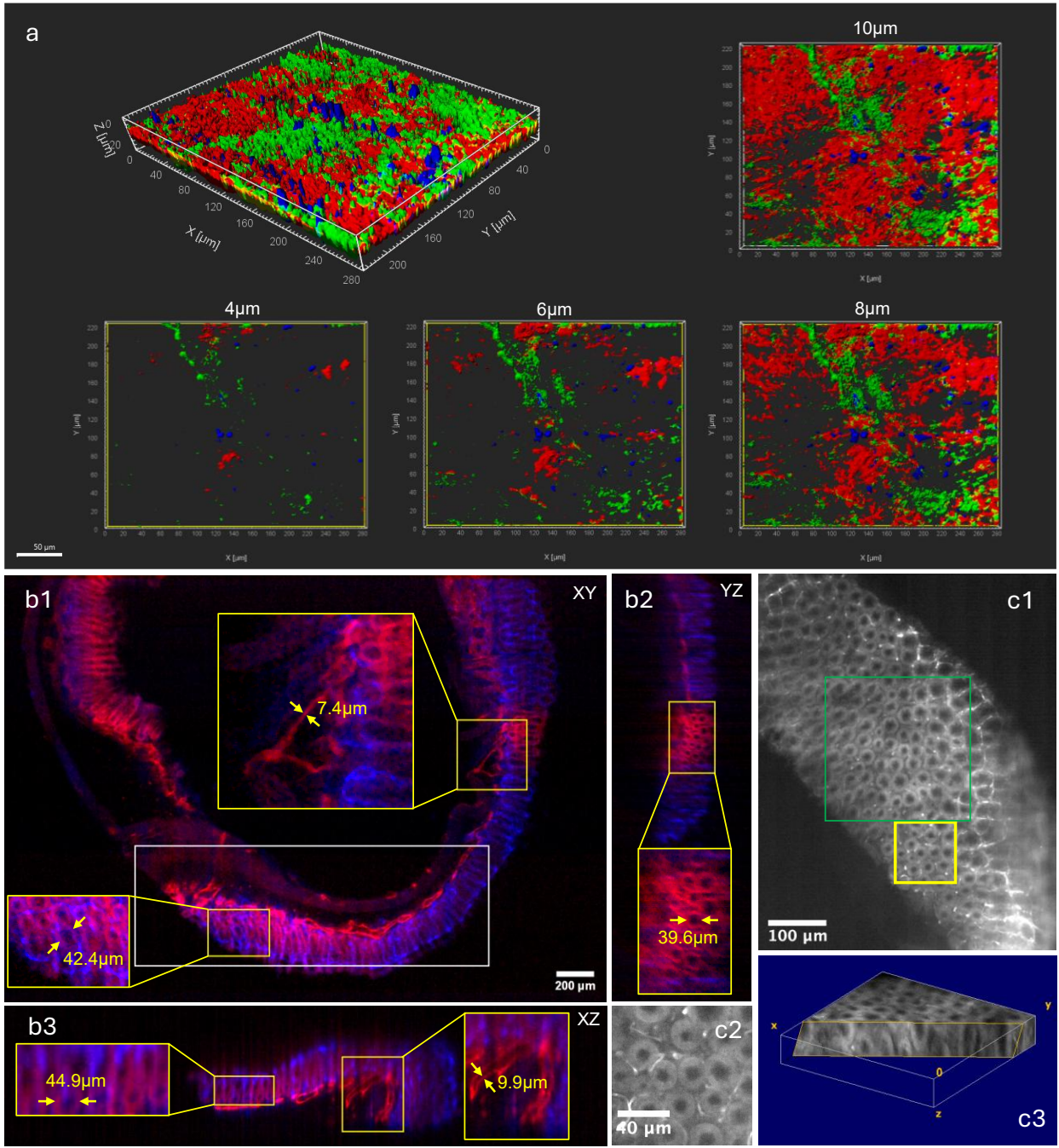

**Figure 3 | Fast scanning of flat biological samples.** (**a**) Fluorescence imaging of mouse lymph nodes. Blue channel: DAPI; Green channel: Cy2 (B cells); Red channel: Cy3 (T cells). (**b**) Low-resolution fluorescence imaging of a mouse

intestine slide. Blue channel: Alexa Fluor® 488; Red channel: SYTOX® Orange.(**c**) High-resolution fluorescence (Alexa Fluor® 488) imaging of the same mouse intestine slide.

The above examples are a manifesto that SLAM is an excellent platform to visualize large, flat biological samples with high-throughput capability and reduced photochemistry. It also provides great flexibility to accommodate various requirements in terms of sample size and resolution.

*2.4 Axial resolution enhancement using a deep learning approach*

Resolution anisotropy is a common issue in microscopy systems. The axial resolution is typically inferior to the transverse resolution especially in case of low NA detection. SLAM inherits the same problem as the NA of the detection objective lens needs be selected to ensure an adequate depth of focus. While conventional methods such as deconvolution and convolutional neural networks (CNN) can restore the axial resolution to some extent, they often produce over-smoothed images or require precise PSF estimation[47-49]. Here we present a novel, diffusion model based approach to computationally enhance the optically affordable axial resolution[50]. The key idea is rather straightforward. The deep neural network learns the structural information from *en face* images, which retains more high-frequency details of the sample. The trained neural network then processes the cross-sectional images of the same sample to restore missing structural information.

Our methodology was validated using a tissue sample slide harvested from the mouse brain hippocampal coronal, which was sectioned by freezing microtome to 30 μm in thickness. The astrocytes within the samples were labeled with the GFAP dye. Original and enhanced images stacks of the sample are compared in **Fig. 4a**, while **Fig. 4b** is a zoomed-in version that includes a single astrocyte. The original volumes (left) were the direct outputs from the SLAM system whilst restored volumes (right) were created from processed images using the diffusion model (see **Methods** for details). Both volume views and orthogonal views were provided for comparing the image qualities. The restored volumes were visually much sharper than the original volumes. In cross-sectional images in both XZ and YZ planes, there were axially elongated and stripe-like structures in the original SLAM images. After diffusion model restoration, however, the structures became more isotropic with finer details. The elliptical shapes of the axon cross-sections were properly delineated only after the processing. Better image quality could also be seen in the transverse images (XY). Nonetheless, the image quality enhancement in this plane was mainly attributed to the elevated contrast rather than resolution as the overlapping structures from neighboring layers were better differentiated. For a semi-quantitative assessment, axial line profiles across the astrocytes and axon were plotted in **Fig. 4c**. One could see that the axon boundaries (membranes) were poorly defined in the original image (dashed lines). After the axial resolution restoration by the diffusion model, sharp peaks (solid lines) appeared with excellent visibility and the full width at half maximum (FWHM) was down to less than 2 μm.

Collectively, these results demonstrated that the diffusion model-based restoration framework effectively improved the axial resolution and structural interpretability of SLAM images. With the help of deep learning-based isotropic restoration, SLAM is capable for volumetric imaging with a high and isotropic resolving power.

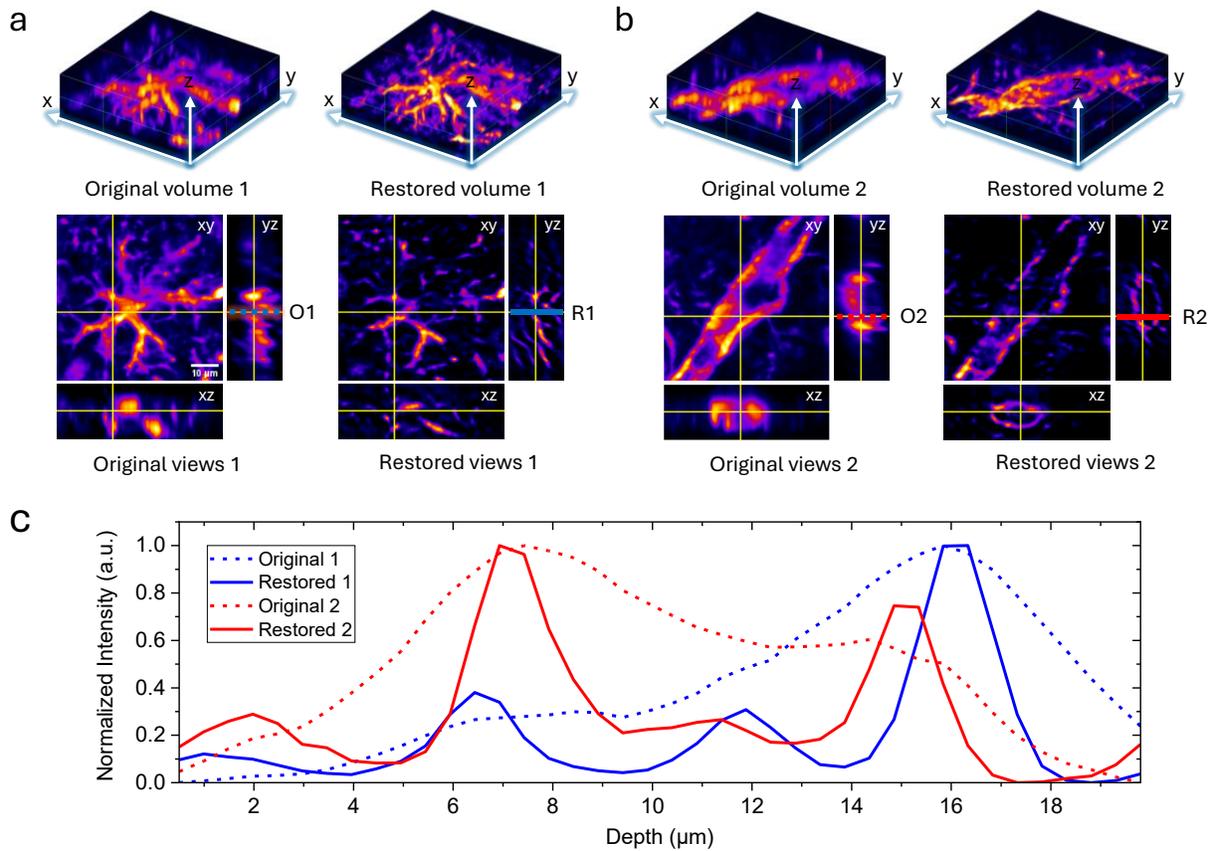

**Figure 4 | Axial resolution enhancement using a deep learning method.** (**a**) SLAM scanning of mouse brain astrocytes tissue slides before and after axial resolution restoration. (**b**) Axon radial structure comparison before and after diffusion model processing. (**c**) Tubular structure comparison before and after diffusion model processing.

## 3   Discussion

The SLAM system design deviates from traditional orthogonal arrangement of the illumination light-sheet and the detection optic axis to achieve many technical advantages including the ease to prepare and handle biological samples and high image acquisition speed ceilings. However, the system configuration must be carefully adjusted to optimally match characteristics of the sample under investigation and to ensure that other image specifications, such as axial and lateral resolutions, can meet specific requirements. It is straightforward to change the detection objective for varying the magnification and numerical aperture, which define the image region in both the transverse and axial directions. Sometimes we insert an adjustable aperture after the DO to reduce the effective detection NA and further increase the depth of field. However, this has to be compromised with the desired lateral and axial resolutions. The overall system axial resolution also depends on the effective thickness of the illumination light-sheet, which is determined by the effective numerical aperture of the illumination objective. We have used 4x (NA 0.1) and 10x (NA 0.3) objectives in this study. Nevertheless, a larger NA could also be considered if even higher axial resolutions are essential.

The non-orthogonal intersection between the illumination point spread function (psf) and the detection psf results in slightly deteriorated optical resolution along the depth direction than conventional light-sheet microscopes if the same illumination NA is used. On the other hand, a relatively high IO NA can typically be used in SLAM as the length of the illumination light-sheet is limited by the sample thickness (or the axial scanning range) rather than the sample width. Therefore, there are many application scenarios in which a slanted light-sheet may lead to better optical resolutions.

The distance between illumination planes in the light-sheet array is around 100 μm in the reported SLAM system with a 4x IO. This parameter is so chosen to balance between the imaging speed and the axial scanning range, in which neighboring light-sheets may generate vertically overlapping emission signals picked up by same camera pixels. For example, the ambiguity depth range is approximately 148 μm for a light-sheet tilt angle of 55.9° at the illumination wavelength of 561 nm. Artifacts may appear in reconstructed images if the sample thickness or the DO depth of field goes beyond this range. In practice, mechanical scanning and image stitching may be still needed for extra thick samples. Alternatively, a single light-sheet could be used when extremely high imaging speed is not necessary, which was the case in our imaging experiments with the FluoTissue 550-μm-thick Mouse Intestine Section.

Computational resolution enhancement has become increasingly popular in the last decade, in which various new deep learning models have been proposed and validated. Light-sheet microscopy is known for its high throughput and imaging speed, but moderate optical resolutions (especially the axial resolution) compared with other techniques such as confocal microscopy. SLAM is no exception. However, we have demonstrated that fundamental optical resolution limitations could be significantly remedied with an advanced deep learning model and a novel training strategy. The diffusion model used in this study showed excellent performance in learning microscopic sample structures and generating high-resolution details from blurred images. However, the computational time for image generation was relatively long. It took about 25 minutes to process a volume of 1296 x 1024 x 40 voxels on a NVIDIA Quadro RTX 6000 GPU. We are interested in exploring other deep-learning models to achieve the similar performance but with a faster image processing speed, which is desirable for real-time image rendering.

The fast scanning mechanism of SLAM is made possible with the combination of light-sheet array generation and a transmission grating for bending the illumination light incident on the sample. As the diffraction angle is wavelength dependent, it is not very straightforward to merge multi-color reconstructed images. Precise and motorized sample stage control and careful calibration is needed for proper image alignment. In addition, the volume phase holographic rating is a relatively expensive component. In the future, we plan to develop a low cost, wavelength independent scanning device that enables the same imaging speed.

**Methods**

*Sample preparation procedures*

**3dpf zebrafish larva with GFP labelled blood vessels.** In this study, we used well-established transgenic lines such as Tg(fli1a:EGFP) or Tg(kdrl:EGFP), in which the EGFP gene was driven by vascular-specific promoters (fli1a or kdrl), leading to selective fluorescence labeling of endothelial cells. These transgenic lines were acquired from zebrafish resource centers (such as the Zebrafish International Resource Center, ZIRC) or through collaboration with research laboratories that maintain these lines. Adult zebrafish were maintained under standard conditions (14/10-hour light/dark cycle at 28.5°C) and were set up for natural spawning to produce embryos. Fertilized embryos were collected and cultured in E3 medium, with 0.003% PTU added at around 12 hours post-fertilization to inhibit pigmentation and enhance imaging clarity. At the appropriate developmental stage (commonly 24–72 hours post-fertilization), embryos were prepared for live or fixed imaging depending on the experimental needs. This process ensures consistent and robust expression of EGFP in the vasculature, facilitating high-quality fluorescence imaging of blood vessels. Prior to imaging, the zebrafish were anesthetized using 0.02% Tricaine and subsequently immobilized in a petri dish using 3.5% methylcellulose, with a thin coverslip forming the dish bottom to facilitate optical access.

As per the guidelines set by the NUS Institutional Animal Care and Use Committee (IACUC), zebrafish at the embryonic or hatched larval stage - up to 5 days post-fertilization and still within the yolk sac stage - are not classified as 'live vertebrate animals.' Therefore, studies involving the above-mentioned specimens were exempt from IACUC review, as the organisms were euthanized by the fifth day post-fertilization.

**Mouse lymph nodes pathology sections.** Branchial lymph nodes were harvested from WT and ApoE-/- mice and embedded in Optimal Cutting Temperature (OCT) (Sakura Finetek). Lymph nodes tissues were cryo-sectioned at 100um on Cryostar NX50 cryostat (Epredia). Sections were dried overnight at room temperature before immunostaining, fixed in acetone and blocked with 0.2% Bovine Serum Albumin (BSA). Primary antibodies Armenian Hamster anti-mouse CD3e (14-0031-85, Clone 145-2C11, eBioscience) (1:100) and Rat anti-mouse B220/CD45R (14-0452-82, Thermo Fisher Scientific) (1:200) were incubated on sections for 3h at room temperature, followed by secondary antibodies Goat anti-Armenian hamster Cy3 (127-165-160, Jackson Laboratories) (1:300) and Donkey anti-rat Cy2 (712-546-150, Jackson Laboratories) (1:300). Sections were counterstained with 4,6-diamidino-2-phenylindole (DAPI) (D1306, Thermo Fisher Scientific) for cell nuclei visualization and mounted with DAKO mounting media for imaging on an Olympus FV3000 confocal microscope. As for staining: Both CD3 (T cells) and B220 (B cells) stain the cell surface (intercellular). Both CD3 (T cells) in Cy3 channel and B220 (B cells) in Cy2 channel stained the cell surfaces.

**Prepared mouse intestine slide.** The sample was ordered from SunJin Lab Co. (https://www.sunjinlab.com/product/fluotissue-550um-mouse-intestine-section/). FluoTissue® prepared slide contains a 550μm fixed tissue section cleared and mounted in RapiClear® 1.52 reagent (Cat. no. RC152001/RC152002). RapiClear® 1.52 is an aqueous tissue clearing reagent not only can highly preserve tissue components but matches the refractive indices of fixed tissue section to that of immersion oil and glass coverslips (1.52nD) to minimize both light scattering and spherical aberrations. The vasculature of the mouse intestine section is labeled with Alexa Fluor® 488, while the cell nuclei are stained using SYTOX® Orange.

**Customized mouse brain slide.** The sample was purchased from Wuhan Servicebio Technology Co., Ltd. The mouse brain hippocampal coronal was sectioned by freezing microtome to 30 μm slices. The mouse brain astrocytes were stained with GFAP. The multicycle TSA-based immunofluorescence staining of frozen sections begins with baking the sections at 37°C for 10-20 minutes, followed by fixation in formaldehyde for 30 minutes and triple PBS washes. Antigen retrieval was performed under specific buffer conditions to preserve tissue integrity, after which endogenous peroxidase activity is blocked using 3% hydrogen peroxide. The tissue was then serum-blocked (10% rabbit serum or 3% BSA depending on antibody source) before adding the first primary antibody and incubating overnight at 4°C in a light-protected humid chamber. After washing, an HRP-conjugated secondary antibody was applied and followed by tyramide signal amplification (TSA). The antibody complex was stripped using a dedicated buffer (RT + 37°C incubations), and the process of blocking, staining, and TSA was repeated for the second and third primary antibodies. After all antibody cycles, DAPI was applied for nuclear counterstaining, and tissue autofluorescence was quenched before final mounting with anti-fade medium.

*Diffusion models*

To address the inherent anisotropic resolution of SLAM, where the axial (Z) resolution is several times worse than the lateral (XY) resolution, we employed a conditional denoising diffusion probabilistic model (DDPM) to enhance the axial resolution[51-53]. Inspired by a previously reported work[54], we trained our model with high-resolution lateral images and the synthetically blurred counterparts. The blurring was simulated by applying Gaussian filtering along the axial direction, where the standard deviation was selected to match the resolution anisotropy observed in the experimental data. Following SR3[55], the blurred and high-resolution pairs were concatenated along the channel dimension at input layer and passed through a U-Net-like architecture to learn the data distribution of the high-resolution image. During inference, the blurred axial image conditioned the generation process to guide the model to generate resolution-enhanced images while maintaining structural consistency. The model predicted a velocity vector as proposed in a previous work[56], which improved the convergence speed and model stability. The model was trained to minimize the mean squared error loss over 1000 time-steps with a cosine noise schedule. To reduce the inference time, only 20 reverse denoising steps were used during sampling. The model was implemented in PyTorch 2.4.0 with CUDA 12.1 support, and trained with the Adam optimizer with a constant learning rate of $1\times10^{-4}$. Training was performed on the NVIDIA Quadro RTX 6000 GPU.

**Disclosures**

National University of Singapore has filed patent applications for the imaging method disclosed in the paper.

**Code, Data, and Materials Availability**

All data in support of the findings of this paper are available from the corresponding author upon reasonable request.

**Acknowledgments**

**Author Contributions:** Conceptualization: NC. Methodology: NC, KL, WC, JZ, JL, SS, ZT, SX, AQ. Investigation: KL, NC, WC. Supervision: NC, AQ. Writing: NC, KL, WC.

**Funding:** This work was supported in part by the following funding sources

Ministry of Education - Singapore MOE2019-T2-2-094 and MOE-000760-01

Science and Technology Project of Jiangsu Province (BZ2022056)

Biomedical and Health Technology Platform, National University of Singapore (Suzhou) Research Institute.